\documentclass[a4paper,11pt]{article}
\pdfoutput=1 

\usepackage{jheppub} 

\usepackage{mathrsfs}
\usepackage{bbm}
\usepackage{bm}
\usepackage{array,longtable}
\usepackage{multirow}
\usepackage{slashed}
\usepackage[utf8]{inputenc}

\allowdisplaybreaks

\title{\boldmath Pseudoscalar meson $P\to \tau (\to \pi \nu_\tau, \rho \nu_\tau, \ell \bar{\nu}_\ell \nu_\tau) \bar{\nu}_\tau$ decays in the Standard Model and beyond}

\author[a]{Quan-Yi Hu}

\affiliation[a]{Department of Physics, Guangxi Normal University, Guilin 541004, Guangxi, China}

\emailAdd{huquanyi@gxnu.edu.cn}

\abstract{In this work, we have conducted a comprehensive and systematic theoretical investigation of the full cascade decays of charged pseudoscalar mesons, specifically $D_s$, $D$, $B$, and $B_c$, into $\tau \nu_\tau$, followed by the subsequent decay of the $\tau$ via its dominant experimentally reconstructible channels: $\tau \to \pi \nu_\tau$, $\tau \to \rho \nu_\tau$, $\tau \to e \bar\nu_e \nu_\tau$, and $\tau \to \mu \bar\nu_\mu \nu_\tau$. Our study is framed within the model-independent low-energy effective field theory approach, which incorporates the most general set of four-fermion operators, including those coupling to right-handed neutrinos. We provide precise Standard Model predictions for differential decay rate as a function of the final-state charged particle energy, develop an innovative and robust methodology for extracting the magnitudes of the new physics couplings using energy moments, and identify and characterize fixed points in the normalized energy distributions. The fixed points are invariant under new physics contributions described by the considered effective Hamiltonian.}

\begin{document} 
\maketitle
\flushbottom

\section{Introduction}
\label{sec:introduction}
Purely leptonic decays of charged pseudoscalar mesons, processes of the type $P^- \to l^- \bar\nu_l$ where $l=e,\mu,\tau$, represent a particularly clean sector within flavor physics. The absence of strongly interacting particles in the final state renders these decays a direct probe of non-perturbative Quantum Chromodynamics (QCD) governing the binding of the constituent quark-antiquark pair within the initial meson. All strong interaction effects are encoded in a single parameter: the pseudoscalar meson decay constant, $f_P$. In the Standard Model (SM), these decays occur via the annihilation of the meson’s quark-antiquark pair into a virtual $W$ boson, which subsequently decays to a charged lepton $l$ and its corresponding neutrino, the decay rate is given by the simple expression (neglecting radiative corrections)
\begin{align}
\label{eq:smdr}
\Gamma(P \to l \bar\nu_l)_\mathrm{SM} = \frac{G_F^2}{8\pi} f_P^2 |V_{q_1 q_2}|^2 m_P m_l^2 \left(1-\frac{m_l^2}{m_P^2}\right)^2,
\end{align}
where $G_F$ is the Fermi coupling constant, $V_{q_1 q_2}$ the relevant Cabibbo-Kobayashi-Maskawa (CKM) matrix element, $m_P$ the meson mass, and $m_l$ the charged lepton mass.

In the SM, purely leptonic decays of pseudoscalar mesons may be subject to CKM suppression and helicity suppression. The pseudoscalar mesons capable of decaying into the final state $\tau \bar\nu_\tau$ include $D_s^-$, $D^-$, $B_c^-$ and $B^-$. Detailed information regarding the squared CKM matrix elements $|V_{q_1 q_2}|^2$, the helicity suppression factors proportional to the squared lepton mass $m_l^2$, and the corresponding phase-space factors $\left(1-m_l^2/m_P^2\right)^2$ for each decay channel is summarized in table~\ref{tab:suppression}. Only the $D_s \to \tau \bar\nu_\tau$ decay is free from both CKM suppression and helicity suppression, rendering it of great experimental and theoretical research interest. Experimentally, the measurements of this decay are mainly carried out by CLEO~\cite{CLEO:2009jky,CLEO:2009lvj,CLEO:2009vke}, BaBar~\cite{BaBar:2010ixw}, Belle~\cite{Belle:2013isi} and BESIII~\cite{BESIII:2016cws,BESIII:2021anh,BESIII:2021wwd,BESIII:2021bdp,BESIII:2023ukh,BESIII:2023fhe,BESIII:2024dvk}. In addition, experimental measurements have also been conducted for the decays of $D \to \tau \bar\nu_\tau$ and $B \to \tau \bar\nu_\tau$. Following the first observation of the $D \to \tau \bar\nu_\tau$ decay at a statistical significance of $5.1\sigma$ by the BESIII collaboration in 2019~\cite{BESIII:2019vhn}, they subsequently reported an improved measurement of this decay in 2024~\cite{BESIII:2024vlt}. Recently, the Belle II collaboration found evidence for the $B \to \tau \bar\nu_\tau$ decay with a significance of $3.0\sigma$~\cite{Belle-II:2025ruy}, including systematic uncertainties. Currently, there are no experiments that have measured the $B_c \to \tau \bar\nu_\tau$ decay. Future Tera-$Z$ machines, such as CEPC~\cite{Zheng:2020ult} and FCC-ee~\cite{Amhis:2021cfy,Zuo:2023dzn}, will be able to directly measure $\mathcal{B}(B_c \to \tau \bar\nu_\tau)$ at the $\mathcal{O}(1\%)$ level.

Due to the extremely short lifetime of the $\tau$ lepton, experimentally, one must reconstruct the $\tau$ lepton through its subsequent decays, typically selecting the following four decay channels — $\tau \to \pi \nu_\tau$~\cite{CLEO:2009lvj,Belle:2013isi,BESIII:2016cws,BESIII:2021anh,BESIII:2023fhe,BESIII:2019vhn,BESIII:2024vlt,Belle-II:2025ruy}, $\tau \to \rho \nu_\tau$~\cite{CLEO:2009vke,BESIII:2021wwd,Belle-II:2025ruy}, $\tau \to e \bar\nu_e \nu_\tau$~\cite{CLEO:2009jky,BaBar:2010ixw,Belle:2013isi,BESIII:2021bdp,Belle-II:2025ruy}, and $\tau \to \mu \bar\nu_\mu \nu_\tau$~\cite{BaBar:2010ixw,Belle:2013isi,BESIII:2023ukh,Belle-II:2025ruy} — each of which contains only a single charged particle in the final state. These modes account for more than 70\% of the total $\tau$ decay width.

\begin{table}[t]
\tabcolsep 0.25in
\renewcommand\arraystretch{1.2}
\begin{center}
\caption{\label{tab:suppression} \small Summary of the CKM suppression and helicity suppression and the corresponding phase-space factor for each decay channel. The squared CKM matrix elements are normalized to $|V_{cs}|^2$, and the helicity suppression and phase-space factors are normalized to the $\tau$ channel.}
\vspace{0.18cm}
{\footnotesize
\begin{tabular}{|c|c|}
\hline
\multicolumn{2}{|l|}{CKM suppression}\\
\hline
\multicolumn{2}{|l|}{$|V_{cs}|^2 : |V_{cd}|^2 : |V_{cb}|^2 : |V_{ub}|^2 \simeq 1 : 5.1\times 10^{-2} : 1.8\times 10^{-3} : 1.5\times 10^{-5}$}\\
\hline
\multicolumn{2}{|l|}{Helicity suppression and phase-space factor $m_l^2 \left(1- m_l^2/m_P^2\right)^2$}\\
\hline
$P$ & $e\bar\nu_e : \mu \bar\nu_\mu : \tau \bar\nu_\tau$ \\
\hline 
$D_s^-$ &  $2.4\times 10^{-6} : 0.1 : 1$\\
\hline 
$D^-$ &  $8.8\times 10^{-6} : 0.4 : 1$\\ 
\hline
$B_c^-$ &  $9.8\times 10^{-8} : 4.2\times 10^{-3} : 1$\\
\hline
$B^-$ &  $1.1\times 10^{-7} : 4.5\times 10^{-3} : 1$\\
\hline 
\end{tabular}
}
\end{center}
\end{table}

The $P \to \tau \bar\nu_\tau$\footnote{Here and throughout the following text, $P$ denotes the $D_s$, $D$, $B_c$, or $B$ meson.} decay is highly sensitive to new physics (NP) beyond the SM, such as models involving a charged Higgs boson~\cite{Branco:2011iw} or leptoquarks~\cite{Buchmuller:1986zs,Dorsner:2016wpm}. At the typical energy scale of the process ($m_b$ for $B_c$ and $B$ mesons, and $m_c$ for $D_s$ and $D$ mesons), these high-scale NP particles decouple. Using a model-independent low-energy effective field theory approach, the most general effective Hamiltonian including right-handed neutrinos can be written as\footnote{Throughout this work, neutrino masses are consistently neglected.}
\begin{align}
	\label{eq:Heff}
	\mathcal{H}_\mathrm{eff}^{q_2 \to q_1 \tau \nu_\tau} = \sqrt{2}G_F V_{q_1 q_2}\sum_{B=L,R} \big[
	&(g_{V,B}^{q_1 q_2} \bar{q}_1\gamma^\mu q_2 + g_{A,B}^{q_1 q_2} \bar{q}_1\gamma^\mu \gamma_5 q_2)\bar{\tau} \gamma_\mu P_B \nu_\tau \nonumber\\
	&+ (g_{S,B}^{q_1 q_2} \bar{q}_1 q_2 + g_{P,B}^{q_1 q_2} \bar{q}_1 \gamma_5 q_2)\bar{\tau} P_B \nu_\tau \nonumber\\[3mm]
	&+ g_{T,B}^{q_1 q_2} (\bar{q}_1\sigma^{\mu\nu} P_B q_2) \bar{\tau}\sigma_{\mu\nu} P_B \nu_\tau \big] + \mathrm{H.c.},
\end{align}
where $\sigma^{\mu\nu} = \frac{i}{2}[\gamma^\mu,\, \gamma^\nu]$ and the chirality projectors $P_{L,R} = (1\mp \gamma_5)/2$. The effects of the high-scale NP particles are all encoded in the Wilson coefficients $g_{i,B}^{q_1 q_2}$, which are defined at the typical energy scale of the process and depend on the flavors of the quark and antiquark in the initial-state meson. In the SM, $g_{V,L}^{q_1 q_2} = -g_{A,L}^{q_1 q_2} = 1$, and all the remaining Wilson coefficients are zero. Generally, they are non-zero and complex.

Within the framework of effective Hamiltonian~\eqref{eq:Heff}, the decay rate of $P \to \tau \bar\nu_\tau$ will be transformed into $\Gamma(P \to l \bar\nu_l)_\mathrm{SM} \left(\left|g_L^{q_1 q_2}\right|^2 + \left|g_R^{q_1 q_2}\right|^2\right)$, where $g_B^{q_1 q_2} \equiv \frac{m_{P}^2}{m_\tau (m_{q_1} + m_{q_2})}g_{P,B}^{q_1 q_2} - g_{A,B}^{q_1 q_2}\;(B=L,R)$. It is easy to see that the contributions of left-handed and right-handed neutrinos appear in the form of the combination $\left|g_L^{q_1 q_2}\right|^2 + \left|g_R^{q_1 q_2}\right|^2$ without any coupling. If the subsequent decay of the $\tau$ lepton is further considered, our previous work~\cite{Hu:2024ozo} has found that the contributions of left-handed and right-handed neutrinos can be well separated by using the distribution of decay rates with respect to the energy of final-state charged particles. 

In this work, we will investigate the differential decay rates of $D_s$, $D$, and $B$ mesons as functions of the energy of the final-state charged particles in the rest frame of the decaying meson, as well as the fixed points that are not affected by any high-scale NP particles. In addition, by introducing energy moments, we present for the first time a method to directly measure $\left|g_L^{q_1 q_2}\right|$ and $\left|g_R^{q_1 q_2}\right|$ using energy moments and decay rates. The rest of this paper is organized as follows. In section \ref{sec:ddrSM}, we present the numerical results for the differential distributions $d\Gamma/dE$ in the SM. In section \ref{sec:moments}, we define the energy moments and their normalized counterparts, and provide the SM predictions for the first-order energy moment and the normalized first-order energy moment for each cascade decay channel. We also investigate a method to extract NP couplings $|g_L^{q_1 q_2}|^2$ and $|g_R^{q_1 q_2}|^2$ using the decay rate (zeroth-order moment) and the first-order energy moment. In section \ref{sec:fixed_point}, we introduce the normalized differential distributions $d\Gamma/(\Gamma dE)$ and analyze their fixed points. Our conclusions are given in section \ref{sec:conclusions}. The general analytical expressions for the differential distributions $d\Gamma/dE$ are provided in the appendix.

\section{Differential distributions $d\Gamma/dE$ in the SM}
\label{sec:ddrSM}

\begin{figure}[t]
	\centering
	\includegraphics[width=0.31\textwidth]{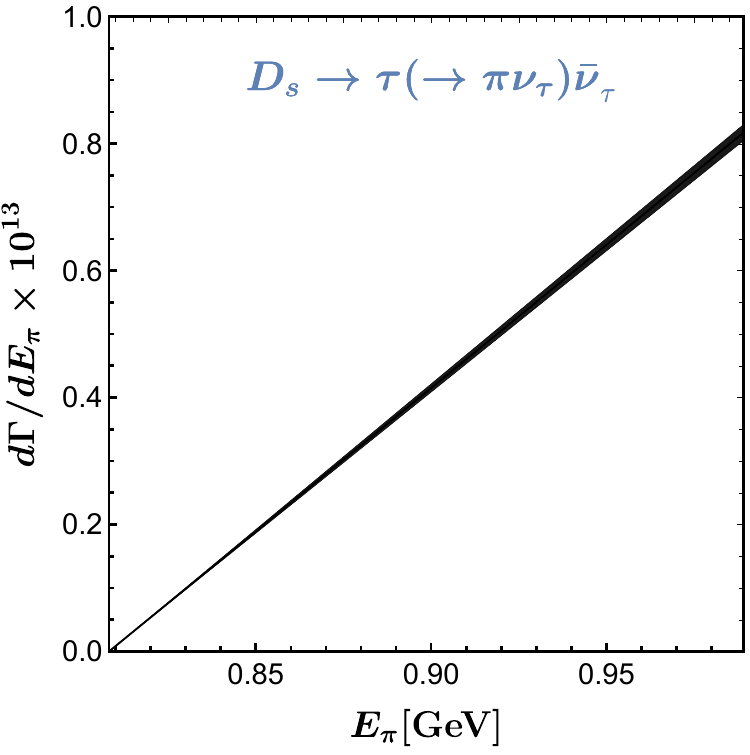}
	\includegraphics[width=0.32\textwidth]{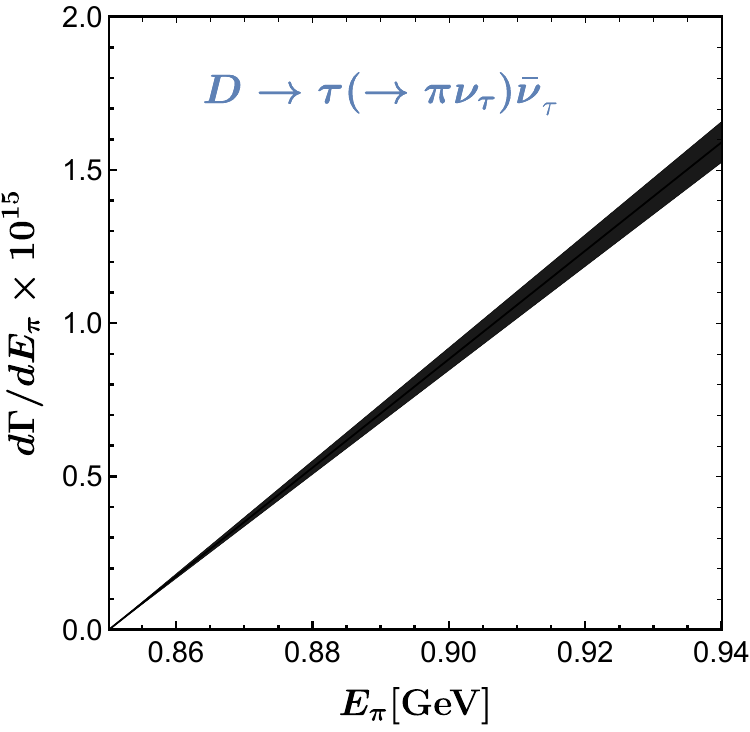}
	\includegraphics[width=0.3\textwidth]{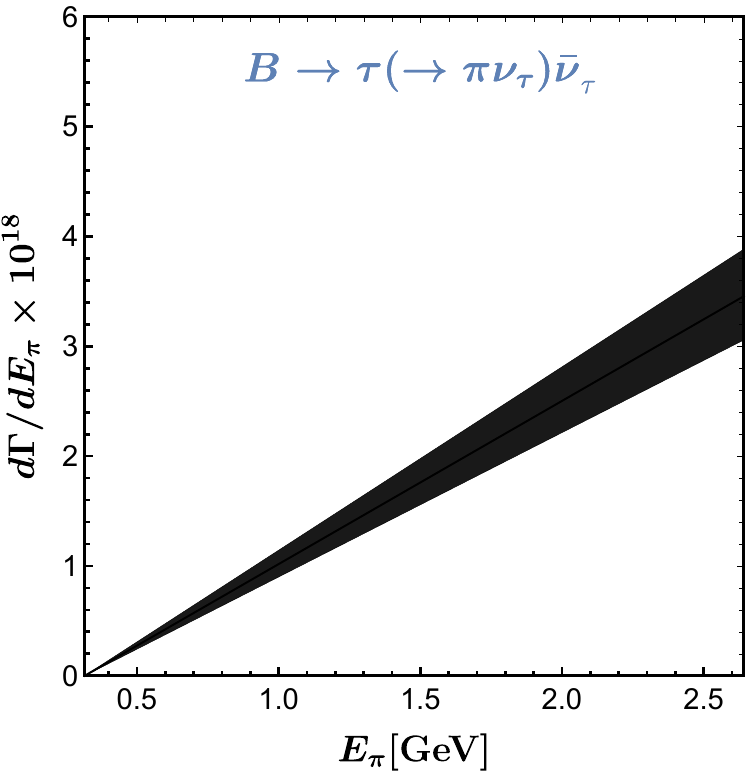}
	\\
	\includegraphics[width=0.315\textwidth]{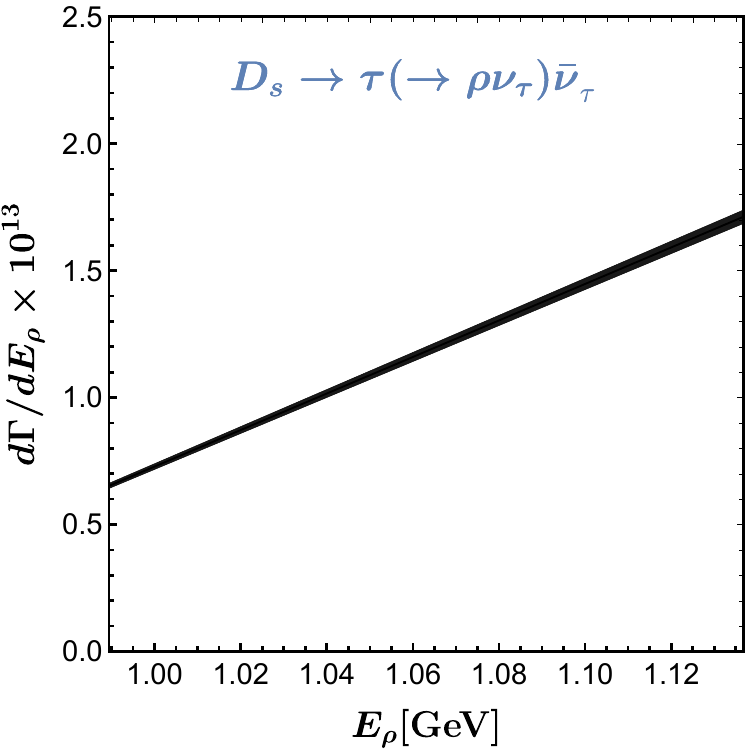}
	\includegraphics[width=0.305\textwidth]{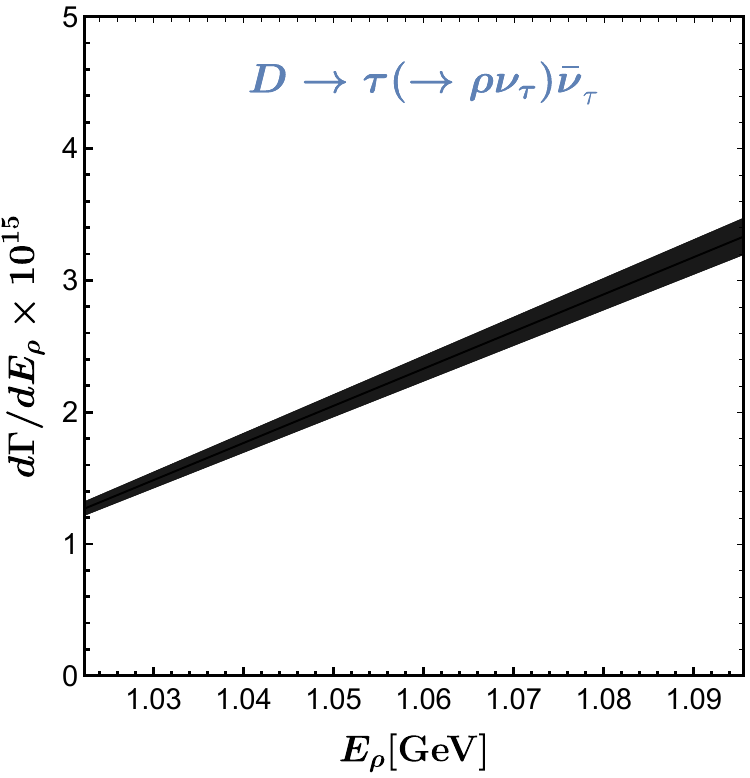}
	\includegraphics[width=0.305\textwidth]{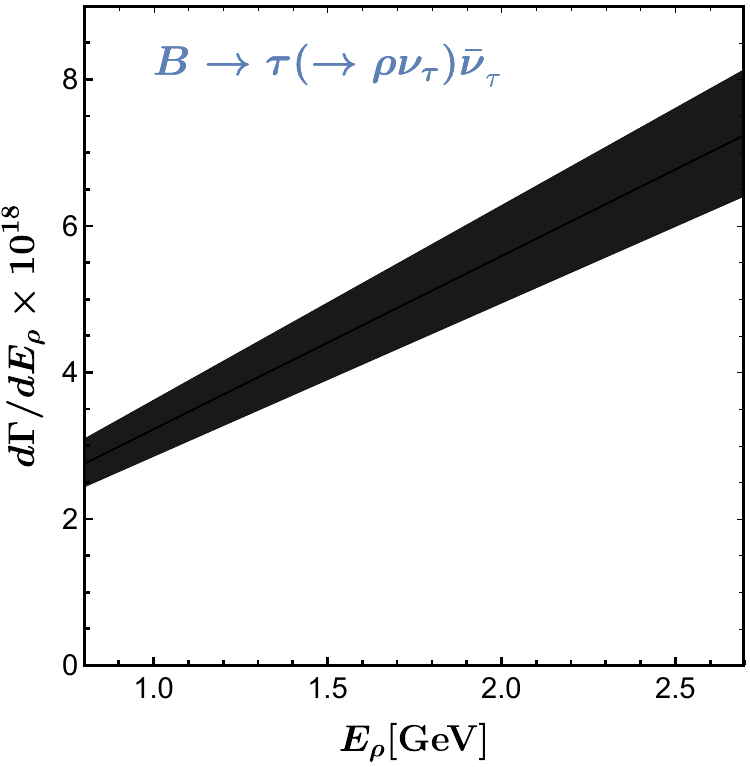}
	\\
	\includegraphics[width=0.315\textwidth]{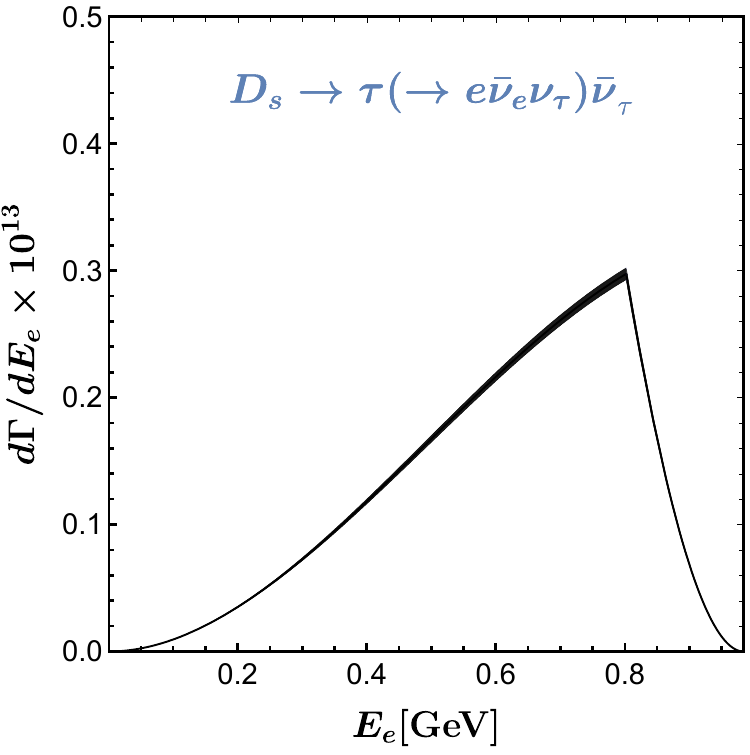}
	\includegraphics[width=0.315\textwidth]{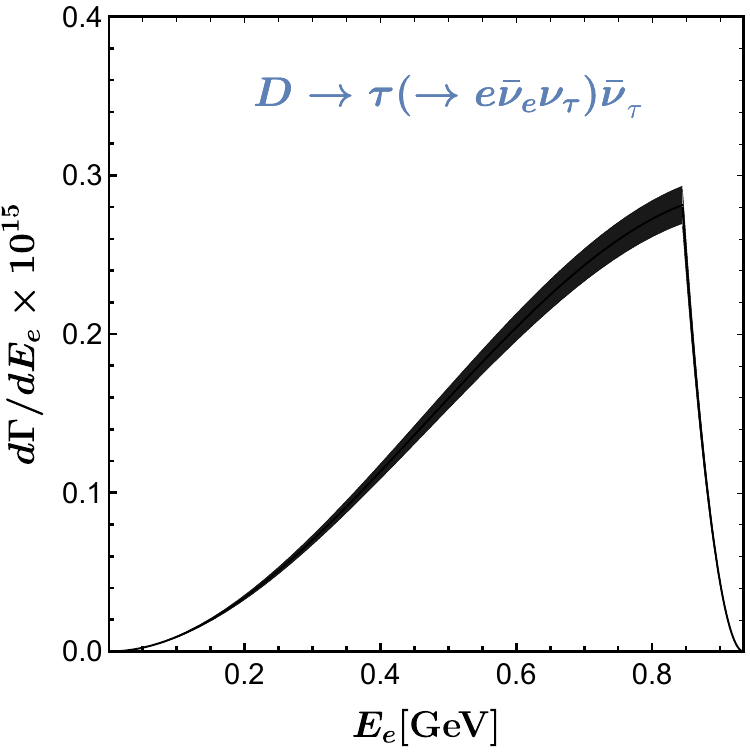}
	\includegraphics[width=0.305\textwidth]{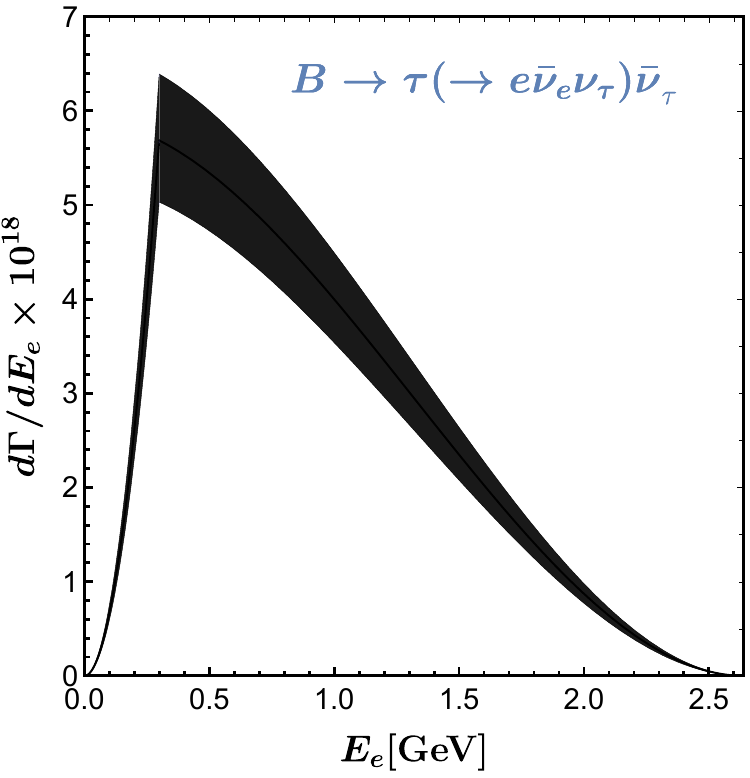}
	\\
	\includegraphics[width=0.315\textwidth]{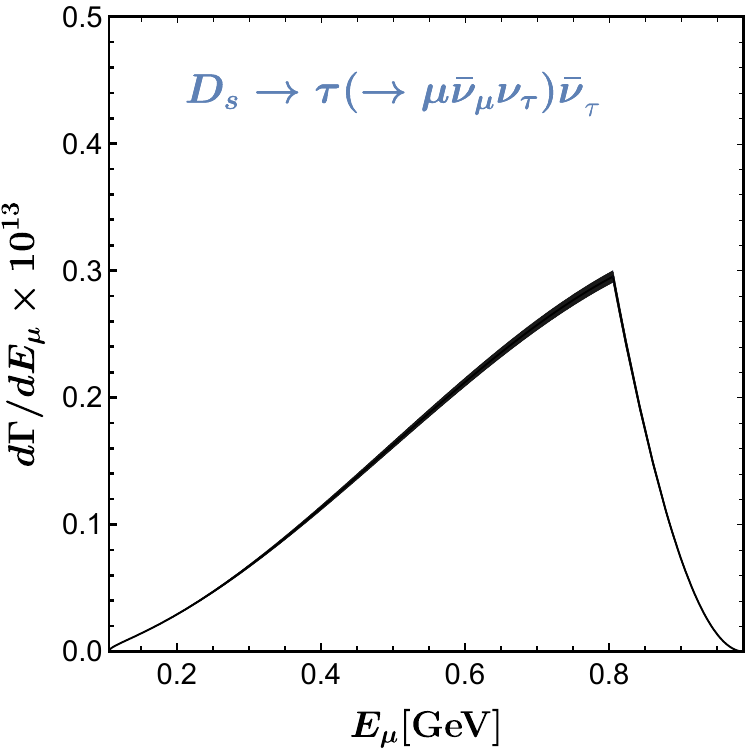}
	\includegraphics[width=0.315\textwidth]{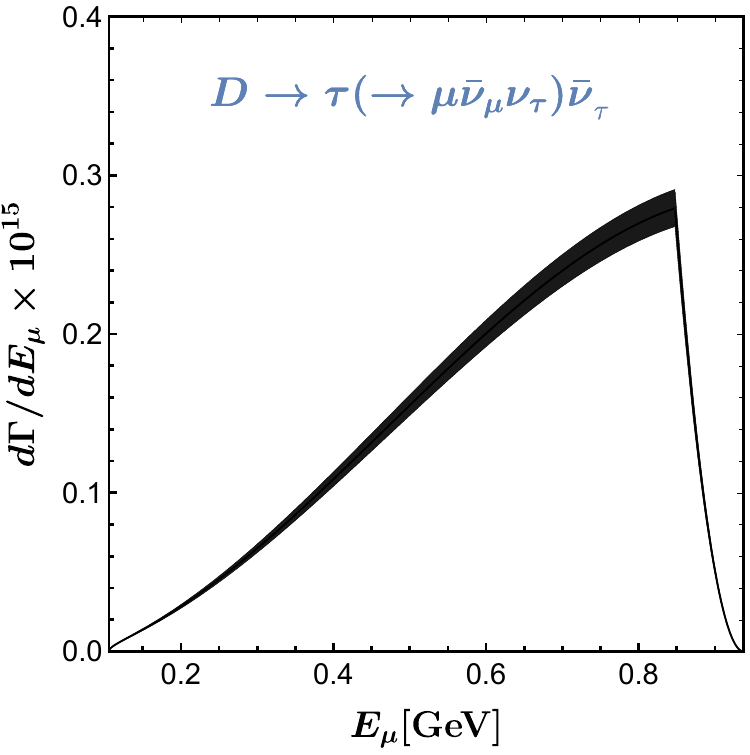}
	\includegraphics[width=0.305\textwidth]{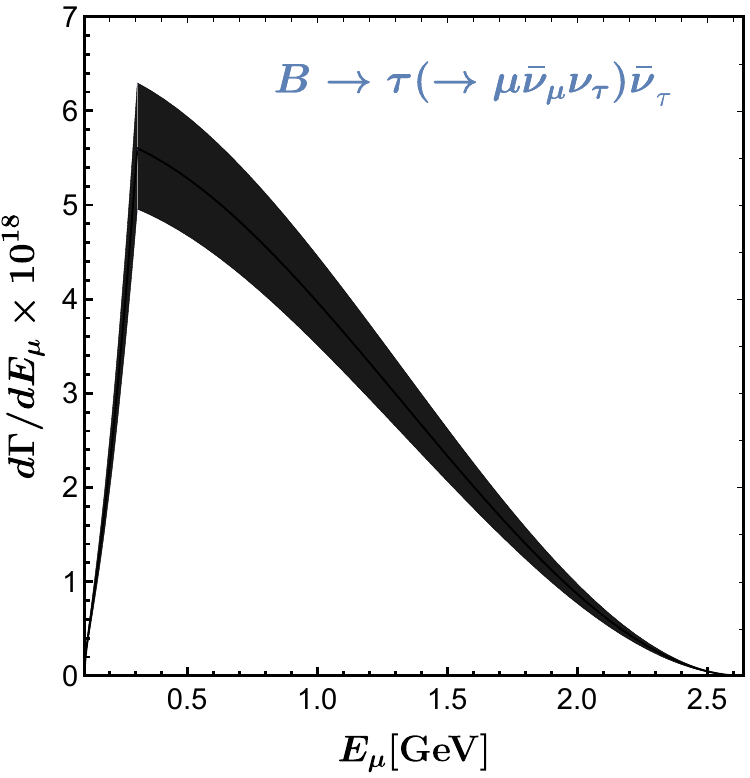}
	\caption{\label{fig:dEsm} \small Numerical results for differential distributions $d\Gamma/dE_a$ ($a=\pi,\rho,e,\mu$) within the SM. From left to right are the results of the cascade decays of the $D_s$, $D$, and $B$ mesons, respectively.}
\end{figure}

In this section, we will present the theoretical predictions within the SM for the purely leptonic decays of $D_s$, $D$, and $B$ mesons into $\tau \bar\nu_\tau$, followed by the subsequent decays of the $\tau$ into $\pi \nu_\tau$, $\rho \nu_\tau$, $e \bar\nu_e \nu_\tau$ and $\mu \bar\nu_\mu \nu_\tau$, specifically focusing on the differential decay rates as a function of the energy of the final-state charged particles (namely $\pi$, $\rho$, $e$, $\mu$). The analytical expressions are provided in the appendix. The decay constants required for the numerical calculations are taken as $f_{D_s} = 249.9(0.4)$~MeV, $f_D = 212.7(0.6)$~MeV, and $f_B = 189.4(1.4)$~MeV~\cite{Bazavov:2017lyh}, all other input parameters are taken from the Particle Data Group~\cite{ParticleDataGroup:2024cfk}.

All numerical results pertaining to this section are compiled in figure~\ref{fig:dEsm}. The dominant sources of uncertainty are the CKM matrix element $V_{q_1q_2}$ and the decay constant $f_P$. The corresponding results for the $B_c$ meson have been reported in ref.~\cite{Hu:2024ozo}. The first column of figure~\ref{fig:dEsm} presents the differential distributions of the cascade decays $D_s\to \tau (\to \pi \nu_\tau, \rho \nu_\tau, e \bar{\nu}_e \nu_\tau, \mu \bar{\nu}_\mu \nu_\tau) \bar{\nu}_\tau$. Compared with the results of the other three mesons, these distributions exhibit the smallest uncertainty and the largest decay rate, rendering them the most promising candidates for the first experimental measurement. The distributions of the decays $D_s\to \tau (\to \ell \bar{\nu}_\ell \nu_\tau) \bar{\nu}_\tau$ and $D\to \tau (\to \ell \bar{\nu}_\ell \nu_\tau) \bar{\nu}_\tau$ are mostly concentrated in the lower energy interval ($m_\ell \leq E_\ell \leq E_\ell^-$), whereas those of the decays $B\to \tau (\to \ell \bar{\nu}_\ell \nu_\tau) \bar{\nu}_\tau$ and $B_c\to \tau (\to \ell \bar{\nu}_\ell \nu_\tau) \bar{\nu}_\tau$~\cite{Hu:2024ozo} are predominantly localized in the higher energy interval ($E_\ell^- \leq E_\ell \leq E_\ell^+$). Experimental measurements of these decays can provide excellent complementarity, enabling us to achieve a comprehensive understanding of the physics across the entire energy range $m_\ell \leq E_\ell \leq E_\ell^+$ in the cascade decays of charged pseudoscalar mesons involving the three-body leptonic subsequent decay of the $\tau$ lepton.

\section{Energy moments and method for measuring NP couplings $g_L^{q_1 q_2}$ and $g_R^{q_1 q_2}$}
\label{sec:moments}

\subsection{Energy moments}
The $n$-th order energy moment of the final-state charged particle $a$ (representing $\pi,\rho,e$, and $\mu$) is defined as
\begin{align}
\label{eq:Mn}
M_a^{(n)} = \int E_a^n \frac{d\Gamma}{dE_a} dE_a.
\end{align}
The corresponding normalized energy moment can be expressed as
\begin{align}
\label{eq:bMn}
\overline{M}_a^{(n)} = \frac{\int E_a^n \frac{d\Gamma}{dE_a} dE_a}{\int \frac{d\Gamma}{dE_a} dE_a}.
\end{align}
Normalized energy moments offer significant advantages both experimentally and theoretically. Experimentally, as both the numerator and the denominator (total decay rate) are measured simultaneously, the normalized energy moments are insensitive to many common systematic errors, which significantly reduces the difficulty and uncertainty of measurements. Theoretically, the CKM matrix element $V_{q_1q_2}$ and the decay constant $f_P$, which are primary sources of uncertainty in theoretical predictions, appear in both the numerator and the denominator and therefore cancel each other out, leading to highly accurate theoretical predictions of the normalized energy moments.

\begin{table}[t]
\tabcolsep 0.15in
\renewcommand\arraystretch{1.2}
\begin{center}
\caption{\label{tab:Ma1} \small Predictions for $M_a^{(1)}$ of all cascade decays within the SM in units of GeV$^2$.}
\vspace{0.18cm}
{\footnotesize
\begin{tabular}{|c|c|c|c|c|}
\hline
       &  $\pi$  &  $\rho$  &  $e$  &  $\mu$\\
\hline 
$D_s$  & $6.87(0.11) \times 10^{-15}$ & $1.87(0.03) \times 10^{-14}$ & $7.51(0.12) \times 10^{-15}$ & $7.44(0.12) \times 10^{-15}$ \\

$D$    & $6.51(0.27) \times 10^{-17}$ & $1.79(0.07) \times 10^{-16}$ & $7.28(0.30) \times 10^{-17}$ & $7.21(0.30) \times 10^{-17}$ \\

$B$    & $7.5(0.9) \times 10^{-18}$ & $1.8(0.2) \times 10^{-17}$ & $6.0(0.7) \times 10^{-18}$ & $6.0(0.7) \times 10^{-18}$ \\

$B_c$  & $6.8(0.9) \times 10^{-15}$ & $1.6(0.2) \times 10^{-14}$ & $5.4(0.7) \times 10^{-15}$ & $5.3(0.7) \times 10^{-15}$ \\
\hline
\end{tabular}
}
\end{center}
\end{table}

\begin{table}[t]
\tabcolsep 0.25in
\renewcommand\arraystretch{1.2}
\begin{center}
\caption{\label{tab:bMa1} \small Predictions for $\overline{M}_a^{(1)}$ of all cascade decays within the SM in units of GeV.}
\vspace{0.18cm}
{\footnotesize
\begin{tabular}{|c|c|c|c|c|}
\hline
	  &  $\pi$  &  $\rho$  &  $e$  &  $\mu$ \\
\hline
$D_s$ & $0.929$ &  $1.07$  & $0.616$ & $0.625$ \\ 
$D$ & $0.910$ &  $1.06$  & $0.618$ & $0.627$ \\ 
$B$ & $1.87$ &  $1.89$  & $0.912$ & $0.926$ \\ 
$B_c$ & $2.18$ &  $2.19$  & $1.04$ & $1.06$ \\ 
\hline
\end{tabular}
}
\end{center}
\end{table}

In the SM, the first-order energy moments $M_a^{(1)}$ and normalized first-order energy moments $\overline{M}_a^{(1)}$ corresponding to each cascade decay channel are presented in tables \ref{tab:Ma1} and \ref{tab:bMa1}, respectively. The decay constant $f_{B_c} = 434(15)$~MeV~\cite{Colquhoun:2015oha} is required when calculating $M_a^{(1)}$ for the cascade decays of $B_c$ meson. Once the complete differential energy spectrum is measured experimentally, the energy moments can be derived. All measurements inconsistent with those in tables \ref{tab:Ma1} and \ref{tab:bMa1}, especially those deviating from $\overline{M}_a^{(1)}$ in table \ref{tab:bMa1}, may indicate the presence of NP.

\subsection{Method for measuring NP couplings $g_L^{q_1 q_2}$ and $g_R^{q_1 q_2}$}

Assuming that the NP beyond the SM can be described by the effective Hamiltonian~\eqref{eq:Heff}. Since the energy moments and decay rates (i.e., the zeroth-order energy moments) exhibit different dependencies on $|g_L^{q_1q_2}|^2$ and $|g_R^{q_1q_2}|^2$, we can arbitrarily select two energy moments of different orders to inversely derive $|g_L^{q_1q_2}|^2$ and $|g_R^{q_1q_2}|^2$. Next, we take the zeroth- and first-order energy moments as examples for illustration.

Using the cascade decay $P \to \tau(\to \pi \nu_\tau)\bar\nu_\tau$, we can extract $|g_L^{q_1q_2}|^2$ and $|g_R^{q_1q_2}|^2$ as follows.
\begin{align}
|g_L^{q_1q_2}|^2 = 8\pi m_P^3 \frac{N_\pi^L}{D_\pi},\;
|g_R^{q_1q_2}|^2 = 8\pi m_P^3 \frac{N_\pi^R}{D_\pi},
\end{align}
where
\begin{align}
N_\pi^L \equiv& +6 m_P m_\tau^2 M_\pi^{(1)} - \left[m_P^2 \left(2m_\pi^2 + m_\tau^2\right) + m_\tau^2 \left(m_\pi^2 + 2m_\tau^2\right)\right] M_\pi^{(0)}, \\
N_\pi^R \equiv& -6 m_P m_\tau^2 M_\pi^{(1)} + \left[m_\tau^2 \left(2m_\pi^2 + m_\tau^2\right) + m_P^2 \left(m_\pi^2 + 2m_\tau^2\right)\right] M_\pi^{(0)}, \\
D_\pi \equiv& \mathcal{B}(\tau \to \pi \nu_\tau) G_F^2 |V_{q_1q_2}|^2 f_P^2 m_\tau^2 \left(m_P^2-m_\tau^2\right)^3 \left(m_\tau^2 - m_\pi^2\right).
\end{align}

Using the cascade decay $P \to \tau(\to \rho \nu_\tau)\bar\nu_\tau$, we have
\begin{align}
|g_L^{q_1q_2}|^2 = 8\pi m_P^3 \frac{N_\rho^L}{D_\rho}, \;
|g_R^{q_1q_2}|^2 = 8\pi m_P^3 \frac{N_\rho^R}{D_\rho}, 
\end{align}
where
\begin{align}
N_\rho^L \equiv &+6 m_P m_\tau^2 (2m_\rho^2 + m_\tau^2) M_\rho^{(1)} \nonumber\\
&- \left[m_P^2 \left(2m_\rho^4 + 6m_\rho^2 m_\tau^2 +m_\tau^4\right) + m_\tau^2 \left( 4m_\rho^4 +3m_\rho^2m_\tau^2 +2m_\tau^4\right)\right] M_\rho^{(0)}, \\
N_\rho^R \equiv &-6 m_P m_\tau^2 (2m_\rho^2 + m_\tau^2) M_\rho^{(1)} \nonumber\\
&+ \left[m_\tau^2 \left(2m_\rho^4 + 6m_\rho^2 m_\tau^2 +m_\tau^4\right) + m_P^2 \left( 4m_\rho^4 +3m_\rho^2m_\tau^2 +2m_\tau^4\right)\right] M_\rho^{(0)}, \\
D_\rho \equiv & \mathcal{B}(\tau \to \rho \nu_\tau) G_F^2 |V_{q_1q_2}|^2 f_P^2 m_\tau^2 (m_P^2 - m_\tau^2)^3 (2m_\rho^4 - 3m_\rho^2m_\tau^2 + m_\tau^4).
\end{align}

Using the cascade decay $P \to \tau(\to \ell \bar\nu_\ell \nu_\tau)\bar\nu_\tau$, we have
\begin{align}
|g_L^{q_1q_2}|^2 = 8\pi \frac{N_\ell^L}{D_\ell}, \;
|g_R^{q_1q_2}|^2 = 8\pi \frac{N_\ell^R}{D_\ell}, 
\end{align}
where
\begin{align}
N_\ell^L \equiv& -60 \left(x^8-8 x^6+24 x^4 \ln x+8 x^2-1\right) M_\ell^{(1)}\nonumber\\
&+ \Big\{\left(x^2-1\right) \Big[7 x^8 \left(2 y^2+1\right)+x^6 \left(7-61 y^2\right)+x^4 \left(119y^2+307\right) \nonumber\\
&-3 x^2 \left(7 y^2+11\right)+9 y^2+12\Big]-120 x^4 \left(3 x^2+y^2+2\right) \ln x\Big\} m_P M_\ell^{(0)}, \\
N_\ell^R \equiv& +60 \left(x^8-8 x^6+24 x^4 \ln x+8 x^2-1\right) M_\ell^{(1)} \nonumber\\
&+\Big\{120 x^4 \left(\left(3 x^2+2\right) y^2+1\right) \ln x-\left(x^2-1\right) \Big[7 x^8 \left(y^2+2\right) +x^6 \left(7 y^2-61\right) \nonumber \\
&+x^4 \left(307 y^2+119\right)-3 x^2 \left(11 y^2+7\right)+12 y^2+9\Big]\Big\} m_P M_\ell^{(0)}, \\
D_\ell \equiv& \mathcal{B}(\tau \to \ell \bar{\nu}_\ell \nu_\tau ) G_F^2 |V_{q_1q_2}|^2 f_P^2 m_P^4 y^2 \left(y^2-1\right)^3 \nonumber\\
&\times\left[7 x^{10}-75 x^8-120 x^6+200 x^4-15
x^2+120 \left(3 x^6+x^4\right) \ln x+3\right].
\end{align}

\begin{table}[t]
\tabcolsep 0.12in
\renewcommand\arraystretch{1.8}
\begin{center}
\caption{\label{tab:gLgR} \small Extracted dimensionless parameters $|g_L^{q_1q_2}|^2$ and $|g_R^{q_1q_2}|^2$ from the cascade decays $P\to \tau (\to \pi \nu_\tau, \rho \nu_\tau, e \bar{\nu}_e \nu_\tau, \mu \bar{\nu}_\mu \nu_\tau) \bar{\nu}_\tau$ for $P = D_s,D,B$, and $B_c$. The extraction is performed using the zeroth- and first-order energy moments $M_a^{(0)}$ (in units of GeV) and $M_a^{(1)}$ (in units of GeV$^2$), respectively, together with the pseudoscalar decay constant $f_P$ (in units of GeV).}
\vspace{0.1cm}
{\scriptsize
\begin{tabular}{|c|c|c|c|c|}
\hline
  & $\tau \to \pi \nu_\tau$ & $\tau \to \rho \nu_\tau$ & $\tau \to e \bar{\nu}_e \nu_\tau$ & $\tau \to \mu \bar{\nu}_\mu \nu_\tau$ \\
\hline 
$\frac{|V_{cs}|^2 f_{D_s}^2}{10^{14}} |g_L^{cs}|^2$ & $1.33 M_\pi^{(1)} - 1.15 M_\pi^{(0)}$ & $1.54 M_\rho^{(1)} - 1.63 M_\rho^{(0)}$ & $1.70 M_e^{(0)} - 2.68 M_e^{(1)}$ & $1.75 M_\mu^{(0)} - 2.72 M_\mu^{(1)}$ \\
$\frac{|V_{cs}|^2 f_{D_s}^2}{10^{14}} |g_R^{cs}|^2$ & $1.24 M_\pi^{(0)} - 1.33 M_\pi^{(1)}$ & $1.66 M_\rho^{(0)} - 1.54 M_\rho^{(1)}$ & $2.68 M_e^{(1)} - 1.65 M_e^{(0)}$ & $2.72 M_\mu^{(1)} - 1.70 M_\mu^{(0)}$ \\
\hline
$\frac{|V_{cd}|^2 f_{D}^2}{10^{15}} |g_L^{cd}|^2$ & $1.03 M_\pi^{(1)} - 0.91 M_\pi^{(0)}$ & $1.20 M_\rho^{(1)} - 1.26 M_\rho^{(0)}$ & $1.30 M_e^{(0)} - 2.08 M_e^{(1)}$ & $1.34 M_\mu^{(0)} - 2.11 M_\mu^{(1)}$ \\
$\frac{|V_{cd}|^2 f_{D}^2}{10^{15}} |g_R^{cd}|^2$ & $0.94 M_\pi^{(0)} - 1.03 M_\pi^{(1)}$ & $1.28 M_\rho^{(0)} - 1.20 M_\rho^{(1)}$ & $2.08 M_e^{(1)} - 1.28 M_e^{(0)}$ & $2.11 M_\mu^{(1)} - 1.32 M_\mu^{(0)}$ \\
\hline
$\frac{|V_{ub}|^2 f_{B}^2}{10^{11}} |g_L^{ub}|^2$ & $1.68 M_\pi^{(1)} - 1.83 M_\pi^{(0)}$ & $1.95 M_\rho^{(1)} - 3.14 M_\rho^{(0)}$ & $3.87 M_e^{(0)} - 3.38 M_e^{(1)}$ & $3.99 M_\mu^{(0)} - 3.43 M_\mu^{(1)}$ \\
$\frac{|V_{ub}|^2 f_{B}^2}{10^{11}} |g_R^{ub}|^2$ & $3.14 M_\pi^{(0)} - 1.68 M_\pi^{(1)}$ & $3.69 M_\rho^{(0)} - 1.95 M_\rho^{(1)}$ & $3.38 M_e^{(1)} - 3.08 M_e^{(0)}$ & $3.43 M_\mu^{(1)} - 3.18 M_\mu^{(0)}$ \\
\hline
$\frac{|V_{cb}|^2 f_{B_c}^2}{10^{11}} |g_L^{cb}|^2$ & $1.07 M_\pi^{(1)} - 1.31 M_\pi^{(0)}$ & $1.24 M_\rho^{(1)} - 2.28 M_\rho^{(0)}$ & $2.85 M_e^{(0)} - 2.14 M_e^{(1)}$ & $2.94 M_\mu^{(0)} - 2.18 M_\mu^{(1)}$ \\
$\frac{|V_{cb}|^2 f_{B_c}^2}{10^{11}} |g_R^{cb}|^2$ & $2.33 M_\pi^{(0)} - 1.07 M_\pi^{(1)}$ & $2.71 M_\rho^{(0)} - 1.24 M_\rho^{(1)}$ & $2.14 M_e^{(1)} - 2.23 M_e^{(0)}$ & $2.18 M_\mu^{(1)} - 2.30 M_\mu^{(0)}$ \\
\hline
\end{tabular}
}
\end{center}
\end{table}

The numerical results for the extracted couplings $|g_L^{q_1q_2}|^2$ and $|g_R^{q_1q_2}|^2$ obtained from the cascade decays of $D_s,D,B$, and $B_c$ mesons are summarized in table~\ref{tab:gLgR}. As an illustrative example, for the decay $D_s \to \tau (\to \pi \nu_\tau) \bar{\nu}_\tau$, the dimensionless parameters $|g_L^{cs}|^2$ and $|g_R^{cs}|^2$ can be determined using the zeroth- and first-order energy moments $M_\pi^{(0)}$ and $M_\pi^{(1)}$ via the expressions 
\begin{align}
|g_L^{cs}|^2 = \frac{1.33 M_\pi^{(1)} - 1.15 M_\pi^{(0)}}{|V_{cs}|^2 f_{D_s}^2} \times 10^{14}, \;
|g_R^{cs}|^2 = \frac{1.24 M_\pi^{(0)} - 1.33 M_\pi^{(1)}}{|V_{cs}|^2 f_{D_s}^2} \times 10^{14}.
\end{align}
In these expressions, the decay constant $f_{D_s}$ is taken in units of GeV, while the energy moments $M_\pi^{(0)}$ and $M_\pi^{(1)}$ are to be substituted with their numerical values in units of GeV and GeV$^2$, respectively. Similar procedures apply to the other meson species and $\tau$ decay channels.

In the SM, $g_L^{q_1q_2} = 1$ and $g_R^{q_1q_2} = 0$, hence the ratio $|g_R^{q_1q_2}|^2/|g_L^{q_1q_2}|^2 =0$. We would like to emphasize that this ratio is independent of the CKM matrix element $V_{q_1q_2}$ and decay constant $f_P$, any experimentally observed deviation from this ratio indicates the existence of NP effects involving right-handed neutrinos.

\section{Normalized distribution $d\Gamma/(\Gamma dE)$ and fixed point}
\label{sec:fixed_point}

\begin{table}[t]
\tabcolsep 0.15in
\renewcommand\arraystretch{1.2}
\begin{center}
\caption{\label{tab:fixed} \small Fixed points $\left(E,\; \frac{1}{\Gamma} \frac{d\Gamma}{dE}\right)$ of the normalized distributions for the cascade decays $P\to \tau (\to \pi \nu_\tau, \rho \nu_\tau, e \bar{\nu}_e \nu_\tau, \mu \bar{\nu}_\mu \nu_\tau) \bar{\nu}_\tau$ ($P = D_s,D,B$, and $B_c$) within the framework of effective Hamiltonian~\eqref{eq:Heff}. Herein, $E$ is expressed in units of GeV and $\frac{1}{\Gamma} \frac{d\Gamma}{dE}$ in units of GeV$^{-1}$.}
\vspace{0.18cm}
{\footnotesize
\begin{tabular}{|c|c|c|c|c|}
\hline
&  $\pi$  &  $\rho$  &  $e$  &  $\mu$\\
\hline 
\multirow{2}{*}{$D_s$}  & \multirow{2}{*}{$(0.8986,5.525)$} & \multirow{2}{*}{$(1.063,6.782)$} & $(0.3314,0.7077)$ & $(0.3431,0.7247)$ \\
 & & &  $(0.8259,1.878)$ & $(0.8294,1.909)$ \\
\hline
\multirow{2}{*}{$D$}    & \multirow{2}{*}{$(0.8951,11.13)$} & \multirow{2}{*}{$(1.059,13.66)$} & $(0.3327,0.7111)$ & $(0.3444,0.7282)$ \\
 & & &  $(0.8511,2.060)$ & $(0.8544,2.095)$ \\
\hline
\multirow{2}{*}{$B$}    & \multirow{2}{*}{$(1.478,0.4299)$} & \multirow{2}{*}{$(1.749,0.5277)$} & $(0.2015,0.4003)$ & $(0.2165,0.4046)$ \\
& & &  $(1.227,0.4923)$ & $(1.233,0.4989)$ \\
\hline
\multirow{2}{*}{$B_c$}  & \multirow{2}{*}{$(1.705,0.3487)$} & \multirow{2}{*}{$(2.017,0.4280)$} & $(0.1747,0.3434)$ & $(0.1911,0.3436)$ \\
& & &  $(1.418,0.4079)$ & $(1.424,0.4133)$ \\
\hline
\end{tabular}
}
\end{center}
\end{table}

The uncertainty induced by the CKM matrix elements $V_{q_1q_2}$ and decay constants $f_P$ can also be eliminated by dividing the differential distributions $d\Gamma/dE$ by the decay rate $\Gamma$. In the absence of right-handed neutrinos, the result will always be consistent with that predicted by the SM. More interestingly, we find that even when contributions from right-handed neutrinos are taken into account, each normalized distribution $d\Gamma/(\Gamma dE)$ has one or two fixed points of its own, which are not affected by any NP effects in effective Hamiltonian \eqref{eq:Heff}. Next, we present the coordinates of the fixed points corresponding to each cascade decay distribution one by one, so as to facilitate experimental verification.

For the cascade decay $P\to \tau(\to h \nu_\tau) \bar\nu_\tau$ ($h=\pi$ or $\rho$), which includes the two-body hadronic decay of the intermediate $\tau$ lepton, there exists only one fixed point corresponding to its normalized distribution, with coordinates~\cite{Hu:2024ozo}
\begin{align}
E_h &= \frac{\left(m_P^2 + m_\tau^2\right) \left(m_h^2 + m_\tau^2\right)}{4 m_P m_\tau^2},\\
\frac{1}{\Gamma} \frac{d\Gamma}{d E_h} &= \frac{2 m_P m_\tau^2}{\left(m_P^2 - m_\tau^2\right) \left(m_\tau^2 - m_h^2\right)}.
\end{align}

For the cascade decay $P\to \tau(\to \ell \bar\nu_\ell \nu_\tau) \bar\nu_\tau$ ($\ell=e$ or $\mu$) involving a three-body leptonic decay of the intermediate $\tau$ lepton, the normalized distribution has two fixed points, one in each of the energy intervals $(m_\ell,\; E_\ell^-)$ and $(E_\ell^-,\; E_\ell^+)$\footnote{Only the numerical result of the fixed point in the energy interval $(E_\ell^-,\; E_\ell^+)$ is presented in ref.~\cite{Hu:2024ozo}, while that in the interval $(m_\ell,\; E_\ell^-)$ is omitted.}. For the electron, its mass can be safely neglected, and the corresponding coordinates of the fixed points are as follows
\begin{align}
E_e = \frac{3 y^2 m_P}{8 \left(1+y^2\right)} , \;
\frac{1}{\Gamma} \frac{d\Gamma}{d E_e} = \frac{9 \left(2 y^4+5 y^2+2\right)}{8
	\left(y^2+1\right)^3 m_P},
\end{align}
in the interval $(m_e=0 \mathrm{GeV},\; E_e^- = \frac{1}{2}ym_\tau)$, and 
\begin{align}
E_e &= \frac{\left(1+y^2+\sqrt{33+18 y^2-15 y^4}\right)
	m_P}{16 \left(2-y^2\right)} , \\
\frac{1}{\Gamma} \frac{d\Gamma}{d E_e} &= \frac{\left(3-2 y^2\right) \left(39 y^4-138
	y^2-\left(1+y^2\right) \sqrt{33+18y^2-15 y^4}+111\right)}{16 \left(1-y^2\right)
	\left(2-y^2\right)^3 m_P},
\end{align}
in the interval $(E_e^- = \frac{1}{2}ym_\tau,\; E_e^+ = \frac{1}{2}m_P)$. Here, the dimensionless parameter $y \equiv \frac{m_\tau}{m_P}$.

Finally, we summarize in table~\ref{tab:fixed} the numerical results for the coordinates of all fixed points of the normalized distributions for each cascade decay channel, under the assumption of arbitrary high-scale NP contributions within the effective Hamiltonian~\eqref{eq:Heff}. Experimental observations of a shift in the fixed-point position would imply the presence of physics beyond the framework outlined in \eqref{eq:Heff}, such as massive right-handed neutrinos or non-integrable light new particles. The massive right-handed neutrinos could modify the phase-space distributions and thereby shift the fixed-point positions. Consequently, experimental tests of the fixed point not only serve to validate the SM, but also act as a powerful probe for exploring a broader landscape of NP scenarios.

\section{Conclusions}
\label{sec:conclusions}

In this work, we conduct a comprehensive investigation of all charged pseudoscalar mesons (namely $D_s$, $D$, $B$, and $B_c$) that can decay to $\tau \bar\nu_\tau$, as well as the four main subsequent decay channels employed in the experimental reconstruction of $\tau$ lepton, namely $\tau \to \pi \nu_\tau$, $\tau \to \rho \nu_\tau$, $\tau \to e \bar\nu_e \nu_\tau$, and $\tau \to \mu \bar\nu_\mu \nu_\tau$. First, we present the numerical results within the SM for the differential distributions $d\Gamma/dE$ of the cascade decays $P\to \tau(\to \pi \nu_\tau, \rho \nu_\tau, e \bar{\nu}_e \nu_\tau, \mu \bar{\nu}_\mu \nu_\tau) \bar\nu_\tau$ of $D_s$, $D$, and $B$ mesons.

Subsequently, as the core theoretical innovation of this work, we propose to measure the NP coupling coefficients $g_L^{q_1 q_2}$ and $g_R^{q_1 q_2}$ by introducing the energy moments. By noting that the total decay rate (the zero-th moment) and higher-order energy moments depend differently on the squares of these couplings, one can solve for them simultaneously. We provided explicit, ready-to-use analytical formulas for extracting $|g_L^{q_1 q_2}|^2$ and $|g_R^{q_1 q_2}|^2$ from measurements of $M_a^{(0)}$ and $M_a^{(1)}$ for each cascade decay channel. This method transforms the differential energy spectrum into a quantitative probe of NP. A key signature of right-handed neutrino involvement would be a non-zero ratio $|g_R^{q_1 q_2}|^2 / |g_L^{q_1 q_2}|^2$, which is independent of $|V_{q_1q_2}|$ and $f_P$.

Finally, we investigate the normalized distribution $d\Gamma/(\Gamma dE)$ and the issue of fixed points therein. We demonstrated that for the two-body hadronic $\tau$ decay channels ($\tau \to \pi \nu_\tau, \rho\nu_\tau$), there exists a single fixed point in the normalized spectrum. For the three-body leptonic channels ($\tau \to \ell\bar\nu_\ell \nu_\tau$), there are two fixed points, one in each of the two characteristic energy intervals. The coordinates of these points (summarized in table~\ref{tab:fixed}) are universal within the framework of the effective Hamiltonian~\eqref{eq:Heff}. They remain unchanged even in the presence of arbitrary high-scale NP contributions. Their experimental verification (such as BESIII, Belle II, CEPC and FCC-ee) would provide a powerful consistency check for the underlying theoretical description, while a discrepancy would point towards physics outside this specific framework, such as right-handed neutrinos with non-negligible masses or light new particles that cannot be integrated out.

\acknowledgments

This work is supported by the National Natural Science Foundation of China under Grant No.~12105002, and the Guangxi Natural Science Foundation under
Grant No.~2023GXNSFBA026270.

\appendix

\section{Analytical expression for the differential distributions $d\Gamma/dE$}
\label{sec:ana_dis}

For the convenience of readers, in this appendix, we present the analytical expressions of the differential decay rates corresponding to the different subsequent decays of the $\tau$ lepton, which are required for the discussions in the main text. The derivation process of these results can be found in ref.~\cite{Hu:2024ozo}. First, we define the following two energy parameters
\begin{align}
\label{eq:rangeEh}
E_a^{-}\equiv \frac{m_{P}^2 m_a^2 + m_\tau^4}{2 m_{P} m_\tau^2}, \;
E_a^{+}\equiv \frac{m_{P}^2 + m_a^2}{2 m_{P}}.
\end{align}
Here, $a$ denotes the charged particle in the final state of the cascade decay, namely $\pi$, $\rho$, $e$, and $\mu$. Next, we will denote the light leptons $e$ and $\mu$ collectively by the symbol $\ell$.

The differential decay rate of cascade decay $P \to \tau(\to \pi \nu_\tau ) \bar{\nu}_\tau $ with respect to the energy of $\pi$ is given by
\begin{align}
\label{eq:dGamm2Epi}
\frac{d\Gamma}{dE_\pi} = &\frac{G_F^2 |V_{q_1q_2}|^2 f_{P}^2 m_\tau^4 \mathcal{B}(\tau \to \pi \nu_\tau)}{4\pi m_{P}^2 (m_\tau^2 - m_\pi^2)^2}
\Bigg[\left( m_{P}^2 - m_\tau^2 \right)\left( m_\tau^2 - m_\pi^2 \right) \left(|g_L^{q_1q_2}|^2 + |g_R^{q_1q_2}|^2\right)  \nonumber\\
&+ 4 m_{P} m_\tau^2 \left(|g_L^{q_1q_2}|^2-|g_R^{q_1q_2}|^2 \right) \left(E_\pi - \frac{E_\pi^{-} + E_\pi^{+}}{2} \right) \Bigg]. 
\end{align}

The corresponding result for the cascade decay $P \to \tau(\to \rho \nu_\tau ) \bar{\nu}_\tau $ is 
\begin{align}
\label{eq:dGamm2Erho}
\frac{d\Gamma}{dE_\rho} = &\frac{G_F^2 |V_{q_1q_2}|^2 f_{P}^2 m_\tau^4 \mathcal{B}(\tau \to \rho \nu_\tau)}{4\pi m_{P}^2 (m_\tau^2 - m_\rho^2)^2 ( m_\tau^2 + 2m_\rho^2 )}
\Bigg[\left( m_{P}^2 - m_\tau^2 \right) \left( m_\tau^4 + m_\tau^2 m_\rho^2 - 2 m_\rho^4 \right) \left(|g_L^{q_1q_2}|^2 + |g_R^{q_1q_2}|^2 \right)  \nonumber\\
&+ 4 m_{P} m_\tau^2 \left(m_\tau^2 - 2 m_\rho^2\right) \left(|g_L^{q_1q_2}|^2 - |g_R^{q_1q_2}|^2 \right) \left(E_\rho - \frac{E_\rho^{-} + E_\rho^{+}}{2} \right)  \Bigg]. 
\end{align}
The energy of $\pi(\rho)$ ranges from $E_{\pi(\rho)}^{-}$ to $E_{\pi(\rho)}^{+}$, i.e., $E_{\pi(\rho)}^- \leq E_{\pi(\rho)} \leq E_{\pi(\rho)}^+$.

In contrast to two-body subsequent $\tau$ decays, for the cascade decay $B_c \to \tau(\to \ell \bar{\nu}_\ell \nu_\tau ) \bar{\nu}_\tau $, the energy of the final-state charged lepton ranges from $m_\ell$ to $E_\ell^+$. The corresponding differential decay rate needs to be presented in two separate intervals.

For $m_\ell \leq E_\ell \leq E_{\ell}^-$,
\begin{align}
\label{eq:dGamm2Elr1}
\frac{d\Gamma}{dE_\ell} = &\frac{G_F^2 |V_{q_1q_2}|^2 f_{P}^2 m_{P}^2 \mathcal{B}(\tau \to \ell \bar{\nu}_\ell \nu_\tau)}{3\pi y^4 (1-8x^2+8x^6-x^8-24x^4 \ln x)}  s \left(1-y^2\right)^2 \nonumber\\
&\times \Big\{\Big[9 \left(x^2+1\right) \left(y^2+1\right) y^2 z+2
x^2 \left(y^4-8 y^2+1\right) y^2 \nonumber\\
&-8 \left(y^4+y^2+1\right) z^2\Big]\left(|g_L^{q_1q_2}|^2 + |g_R^{q_1q_2}|^2\right) \nonumber\\
&-  \left(1-y^2\right) \left[x^2 y^2 \left(2 y^2+9 z+2\right)+z \left(y^2 (3-8
z)-8 z\right)\right] \left(|g_L^{q_1q_2}|^2 - |g_R^{q_1q_2}|^2\right) \Big\}.
\end{align}

For $E_{\ell}^- \leq E_\ell \leq E_{\ell}^+$,
\begin{align}
	\label{eq:dGamm2Elr2}
	\frac{d\Gamma}{dE_\ell} = &\frac{G_F^2 |V_{q_1q_2}|^2 f_{P}^2 m_{P}^2 \mathcal{B}(\tau \to \ell \bar{\nu}_\ell \nu_\tau )}{12\pi y^4 (1-8x^2+8x^6-x^8-24x^4 \ln x)}\nonumber\\ 
	&\times \Bigg\{
	\left(y^2-1\right) \Big[x^2 \Big(y^8 (4 s+12 z-9)+9 y^6 (2 (s+z) (z-2)+3) \nonumber\\
	&+9 y^4 (4 s-4 z-1)+2 y^2 (3 z (3 z+2)-s (9 z+2))\Big) \nonumber\\
	&-y^6 (2 z (8 z-9)
	(s+z)+5)+18 y^2 z (z-s)+16 z^2 (s-z) \nonumber\\
	&-5 x^6 y^6-9 x^4 y^4 \left(y^4-3 y^2+1\right)\Big] \left(|g_L^{q_1q_2}|^2 + |g_R^{q_1q_2}|^2\right)  \nonumber\\
	&+ \Big[2 s \left(y^2-1\right)^3 \left(x^2 y^2 \left(2 y^2+9 z+2\right)+z \left(3
	y^2-8 \left(y^2+1\right) z\right)\right) \nonumber\\
	&+6 y^2 z^2 \left(3 x^2+1\right) \left(y^2+1\right)
	\left(y^4-4 y^2+1\right) \nonumber\\
	&+12 x^2 y^2 z \left(\left(4 x^2+6\right) y^4+y^8-2 y^6-2
	y^2+1\right) \nonumber\\
	&+y^4 \left( \left(1-5 x^6 \right) y^2 \left(y^2+1\right)- 3 x^2(3 x^2 + 1) \left(y^6+1\right)\right) \nonumber\\
	&-16 \left(y^8-2 y^6-2 y^2+1\right) z^3\Big] \left(|g_L^{q_1q_2}|^2 - |g_R^{q_1q_2}|^2\right)
	\Bigg\}.
\end{align}
Here, the dimensionless parameters are defined as follows
\begin{align}
	x \equiv \frac{m_\ell}{m_\tau},\
	y \equiv \frac{m_\tau}{m_P}, \
	z \equiv \frac{E_\ell}{m_P},\
	s \equiv \sqrt{z^2 - x^2 y^2}.
\end{align}

Setting $g_L^{q_1q_2} = 1$ and $g_R^{q_1q_2} = 0$ yields the differential distributions in the SM.

\bibliographystyle{JHEP}
\bibliography{ref}

\end{document}